\documentclass[doublecol]{epl2}

\usepackage{latexsym,epsfig,graphicx}
\usepackage{dcolumn}
\usepackage{bm}
\usepackage[colorlinks,urlcolor=blue,citecolor=blue]{hyperref}
\usepackage{comment}

\title{Physics of hollow Bose-Einstein condensates}
\shorttitle{Physics of hollow Bose-Einstein condensates}

\author{Karmela Padavi\'{c}\inst{1} \and
Kuei Sun\inst{2}\thanks{kuei.sun@utdallas.edu} \and Courtney
Lannert\inst{3,4}\thanks{clannert@smith.edu} \and Smitha
Vishveshwara\inst{1}\thanks{smivish@illinois.edu} }

\shortauthor{Karmela Padavi\'{c}, Kuei Sun, Courtney Lannert, and
Smitha Vishveshwara}

\institute{
  \inst{1} Department of Physics, University of Illinois at
Urbana-Champaign - Urbana, Illinois 61801-3080, USA\\
  \inst{2} Department of Physics, The University of Texas at
Dallas - Richardson, Texas 75080-3021, USA \\
  \inst{3} Department of Physics, Smith College - Northampton,
Massachusetts 01063, USA \\
  \inst{4} Department of Physics, University of Massachusetts - Amherst, Massachusetts 01003-9300, USA }


\abstract{Bose-Einstein condensate shells, while occurring in
ultracold systems of coexisting phases and potentially within
neutron stars, have yet to be realized in isolation on Earth due
to the experimental challenge of overcoming gravitational sag.
Motivated by the expected realization of hollow condensates by the
space-based Cold Atomic Laboratory in microgravity conditions, we
study a spherical condensate undergoing a topological change from
a filled sphere to a hollow shell. We argue that the collective
modes of the system show marked and robust signatures of this
hollowing transition accompanied by the appearance of a new
boundary. In particular, we demonstrate that the frequency
spectrum of the breathing modes shows a pronounced depression as
it evolves from the filled sphere limit to the hollowing
transition. Furthermore, when the center of the system becomes
hollow surface modes show a global restructuring of their spectrum
due to the availability of a new, inner, surface for supporting
density distortions. We pinpoint universal features of this
topological transition as well as analyse the spectral evolution
of collective modes in the experimentally relevant case of a
bubble-trap.}

\begin{document}

\maketitle

Quantum matter, when subject to transitions of a topological
nature, undergoes fundamental changes in its
properties~\cite{Thouless1973Haldane1983}. Such transitions
involve singular deformations of the underlying space inhabited by
the system, be it real or abstract. For instance, the ripping
action required to convert a sphere into a torus. In topological
materials, which have recently gained prominent attention, matter
can transform from being a trivial insulator to one having gapless
surface states. The topological nature of the transition can be
pinpointed through the calculation or measurement of an abstract
Berry phase type global invariant associated with non-trivial
winding, for instance, in momentum
space~\cite{Hasan2010Qi2011Chiu2016}. Cold atomic systems, given
their spectacular trapping and tuning capabilities, not only
enable measuring such topological
invariants~\cite{Atala2013,Delplace2011,Price2013}, they offer a
much more direct, real space version of a topological transition
through purely changing physical geometry. As a pioneering
instance, the realization of toroidal Bose-Einstein condensates
(BECs)~\cite{Gupta2005, Ramanathan2011} corresponds to a
topological structure characterized by a homotopy group that is
not equivalent to that of a disk. Here we explore salient features
associated with the hollowing out of a spherically filled BEC and
subsequent formation of a closed, hollow shell. The filled and
hollow BECs are topologically inequivalent in that they correspond
to different second homotopy groups. In other words, unlike for
the filled spherically symmetric BEC, a spherical surface within
the hollow BEC that surrounds its center cannot be continuously
deformed into a point. In this sense, the hollowing of a
condensate corresponds to a topological transition.

The study of hollow condensates is particularly germane now in
light of the scheduled launch of the Cold Atomic Laboratory
(CAL)~\cite{CAL} later this year aimed to investigate ultracold
quantum gases aboard the International Space Station. One planned
experiment on board involves the first creation of a hollow shell
BEC~\cite{Lundblad,Hollow} using an rf-dressed ``bubble trap''
potential~\cite{Zobay2001}. While BECs have been produced in a
host of interesting geometries
~\cite{Gorlitz2001,Greiner2001,Dettemer2001,Hechenblaikner2005,Smith2005,Ramanathan2011,
Gupta2005,Gaunt2013,Shin2004,Hofferberth2006,Smerzi1997},
gravitational sag has prevented the realization of BEC shells on
Earth~\cite{Colombe2004,Merloti2013}. Thus, microgravity
environments, as expected to be produced in CAL and successfully
demonstrated in the context of BECs in the ZARM drop tower in
Germany~\cite{ZARM} and in that of normal fluid shells during
space shuttle launches~\cite{Wang94}, are necessary for the
realization of hollow condensates. Shell-shaped BECs would be a
test bed for quantum fluids in this topology, which naturally
occurs in a number of diverse situations from laboratory-based
micron scales to astronomical scales. In cold atomic systems,
condensate shells are expected in Bose-Fermi
mixtures~\cite{Molmer1998,Ospelkaus2006,Schaeybroeck2009} and
optical-lattice ``wedding-cake''
structures~\cite{Batrouni2002,DeMarco2005,Campbell2006} where
Mott-insulating regions effectively trap superfluid
layers~\cite{Barankov2007Sun2009}. In neutron stars, extremely
high mass densities render ambient {\it relative} temperatures low
enough for realizing macroscopic quantum states of matter,
possibly giving rise to shells of differing states, some
corresponding to superconductors and BECs of subatomic
particles~\cite{Weber2004,Pethick2015}.

Collective excitations offer an excellent probe of shape, boundary
constraints, and topology, be they in solids, classical liquids,
or quantum fluids. For instance, one ``hears" the shape of a drum
through its normal-mode oscillations~\cite{Kac1966,Giraud2010}. In
trapped condensates, the low-lying excitations---their collective
modes---were among the first phenomena to be studied after the
initial realization of
BECs~\cite{Jin1996,Mewes1996,Edwards1996,Stringari1996,Castin1996,Stamper-Kurn1998,Chevy2002,Fort2003,Yang2009,Haller2009,Pollack2010,Kuwamoto2012}.
Relevant to our studies, the topological change embodied in the
emergence of an additional surface is significant for various
physical systems. For example, for ships entering shallow water
from the deep seas, the structure of water waves changes when the
ocean floor emerges as a relevant boundary~\cite{Lighthill1978}.
In the aforementioned space shuttle experiments~\cite{Wang94} on
normal fluid shells, sloshing modes show marked signatures of
inner and outer boundaries, akin to those predicted in preliminary
theoretical work on BEC shells~\cite{Lannert2007}. Studies of
two-dimensional annular BECs~\cite{Stringari2006,Perrin2012} have
revealed mode spectra that can distinguish an inner boundary in a
non-destructive fashion. Such a collective mode analysis becomes
even more pertinent for three-dimensional hollow condensates where
direct imaging may not discern the presences of a hollow region.
Here, we show that the collective mode spectrum presents a natural
and powerful way to observe the topological change from filled to
hollow in BECs, pinpointing universal features as well as making
concrete predictions in experimentally relevant settings.

\begin{figure}[t]
\centering
\includegraphics[width=8.6cm]{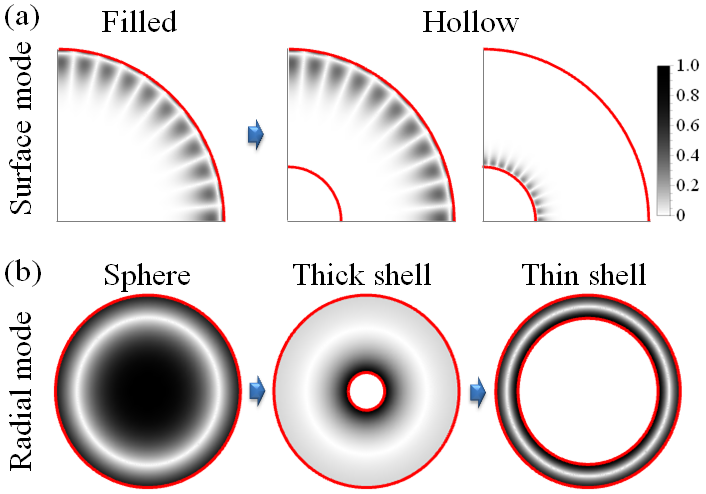}
\caption{(Color online) (a) Density variations of surface
collective modes in the filled sphere (left), and hollow shell
(middle and right), showing the bifurcation of modes confined to
the inner and outer surfaces after the system hollows. (b) Density
variation of breathing modes through the transition from filled
sphere (left) to thin shell (right), showing the localization of
mode amplitude to the (hollowing) center of the system during the
transition. Surface (radial) modes profiles are presented in a
quadrant (full) section of sphere and normalized as in the bar
graph. The red solid curves indicate the (Thomas-Fermi) boundaries
of the system. } \label{fig:Fig1}
\end{figure}

As we discuss in this Letter, the evolution of the collective mode
frequencies as a function of a tuning parameter signals the
appearance of a new boundary when a condensate undergoes the
topological change of hollowing. The effect of the new inner
boundary becomes manifest both in collective modes whose
oscillations are primarily along the surface of the condensate
(surface modes) as well as primarily transverse to it (breathing
modes). In the former case of surface modes
[Fig.~\ref{fig:Fig1}(a)], such as high angular momentum modes in a
sphere~\cite{AlKhawaja1999}, density distortions exponentially
decay into the bulk. When the new boundary appears, for any given
surface mode, radial distortions are redistributed between the two
boundaries. The frequency spectrum shows a sharp jump
corresponding to this redistribution---each boundary surface hosts
fewer nodes of oscillation so the energy associated with a given
oscillatory distortion is lowered.

For the latter case of  transverse modes [Fig.~\ref{fig:Fig1}(b)],
of which the simplest is the spherically symmetric breathing
mode~\cite{Lobser2015Straatsma2016,Lannert2007}, the collective
mode spectrum shows a distinct dip in frequency when the inner
surface is created. Breathing modes localize near the new boundary
as the condensate begins to hollow. Density and stiffness of the
condensate in the center of the system as it hollows are very low,
causing an unusually low oscillation frequency and an associated
dip in the mode spectrum. We explicitly demonstrate these effects
in the simplest case of spherical symmetry; our results are
summarized in Figs.~\ref{fig:Fig2} and \ref{fig:Fig4}.

As is common and effective for collective mode analyses in BECs,
we employ the hydrodynamic approach, which models excitations with
relatively small, smooth deviations from the equilibrium density
of the condensate. (Oscillations of certain low lying modes have
been obtained by a variational approach in the
literature~\cite{Lannert2007} as well.) The collective eigenmode
frequency $\omega$ is then related to density fluctuations in
space and time via $\delta n(\textbf{r},t)=\delta
n(\textbf{r})e^{-i\omega t}$~\cite{Stringari1996,Pethick2008}. The
eigenvalue equation for the collective modes and their frequencies
is given by
\begin{eqnarray}\label{eq:delta_n_no_TF2}
-\frac{mS_l^2}{U}\omega^2\delta n= (\nabla n_{ \rm{eq}})\cdot
(\nabla\delta n)+n_{ \rm{eq}}\nabla^2\delta n,
\end{eqnarray}
where the trapping potential enters through the equilibrium
density profile, $n_{\rm{eq}}(\mathbf{r})$, $m$ is the particle
mass, $S_l$ is the characteristic length for the trapping
potential (in terms of which all lengths are measured below), and
$U=8\pi a_s/S_l$ is the interaction strength (proportional to the
two-body scattering length $a_s$)~\cite{Pethick2008}.
 In the presence of spherical symmetry the
density fluctuation can be expanded in terms of spherical
harmonics as $\delta n(\textbf{r})=D(r)Y_{\ell,m_{\ell}}(\theta,
\phi)$ so that Eq.~(\ref{eq:delta_n_no_TF2}) reduces to

\begin{eqnarray}\label{eq:eigenproblem}
\frac{mS_l^2}{U}{\omega ^2 r^2}D = - \partial_r \left( r^2
n_{\rm{eq}} \partial_r D \right )+\ell(\ell+1)n_{\rm{eq}}D.
\end{eqnarray}

The collective modes in a spherically symmetric system can be
characterized by quantum numbers $\nu, \ell, m_{\ell}$
corresponding to the number of radial and angular nodes,
respectively.

In our solution of the hydrodynamic equations below, we initially
make the simplifying assumption that the condensate is strongly
interacting in the sense that $Na_s/S_l \gg 1$ and that the
associated Thomas-Fermi approximation holds. The Thomas-Fermi form
for the equilibrium density depends on the trapping potential and
the chemical potential, $\mu$ (determined by the net number of
particles, $N$, in the condensate): $n_{\rm{eq}}(\mathbf{r})=
[\mu-V(\mathbf{r})]/U$.  This approximation models a condensate
having sharp boundaries identified by $V(\mathbf{r})=\mu$, while
in reality, the condensate density decreases gradually at its
boundaries. We therefore use the Thomas-Fermi approximation to
pinpoint signatures of hollowing, and extend our results to more
physically realistic scenarios by treating the boundary of the
condensate more accurately through numerics.

While we focus on universal collective mode features, in tandem,
as a concrete example, we analyze the specific case of the
spherical bubble trap, originally proposed by Zobay and
Garraway~\cite{Zobay2001}. This trapping potential allows for
smooth tuning of the condensate shape between the filled sphere
and hollow shell topology and has the form

\begin{equation}\label{eq:bubble}
V_{\rm{bubble}}(\mathbf{r})=m\omega_0^2 S_l^2
\sqrt{(r^2-\Delta)^2/4+\Omega^2}.
\end{equation}

Here $\Delta$ and $\Omega$ are experimental parameters that can be
precisely controlled. The minimum of this potential is found at
$r_0=\sqrt{\Delta}$. The filled sphere BEC geometry is realized
when $\Delta=\Omega=0$ since for this choice of parameters the
bubble trap reduces to the harmonic trap:
$V_{\rm{0}}(\mathbf{r})=\frac{1}{2}m\omega_0^2 S_l^2 r^2$. For
large $\Delta$ (when $r_0$ is much larger than the thickness of
the condensate shell), the bubble trap potential is approximated
near its minimum by a radially-shifted harmonic trap
$V_{\rm{sh}}(\mathbf{r})=\frac{1}{2}m\omega_{\rm{sh}}^2 S_l^2
(r-r_0)^2$ having frequency $\omega_{\rm{sh}} = \omega_0
\sqrt{\Delta/\Omega}$. Below, we set $\Delta=\Omega$ for
simplicity. Consequently, we see that increasing or decreasing the
trap parameter $\Delta$, at a constant chemical potential, results
in a deformation between the filled and hollow condensate
geometries.

\section{Evolution of surface modes}
The effect of hollowing on surface modes of a spherical condensate
can be gleaned from the eigenvalue equation for the collective
mode frequencies, Eq.~(\ref{eq:eigenproblem}), in the limit of
large angular momentum, $\ell  \gg 1$. The centrifugal term
$\ell(\ell+1) n_{\rm{eq}}$ dominates in this limit and is
minimized at the condensate's outer surface (due to $n_{\rm{eq}}
\sim 0$), and also the inner one in the hollow case, thus causing
localization of large-$\ell$ density deviations to these surfaces.
At the transition between a filled and a hollow condensate,  the
doubling of the surfaces available for excitations creates a
redistribution of radial nodes---some of the nodes can localize in
the vicinity of the new, inner surface. Consequently, since fewer
nodes are compressed in each surface region, the frequency of any
single mode (indexed by the total number of radial nodes $\nu$) is
reduced across the hollowing transition. To provide an intuitive
physical picture, we note that a similar situation is found in the
energy states of a quantum double-well potential as the minima are
brought into alignment. At the point at which the minimum
potential values are equal, the energy eigenvalue corresponding to
$\nu$ nodes jumps to corresponding to two degenerate states having
$2\nu$ and $2\nu +1$ nodes (for instance, states having $\nu=2$
and $\nu=3$ radial nodes become nearly degenerate with the $\nu=1$
state.) Here, the centrifugal term in Eq.~(\ref{eq:eigenproblem})
plays the role of the potential and, as we show below, the
experimentally relevant bubble trap even supports the degeneracy
exhibited by the simple double-well analogy.

More concretely, focusing on the two surfaces, the Thomas-Fermi
approximation for the bubble trap identifies the inner and outer
boundaries at the radii $r=R_{\rm{in}}=\sqrt{2\Delta-R^2}$ and
$r=R_{\rm{out}}=R$, respectively. Linearizing the trapping
potential at these boundaries gives the Thomas-Fermi equilibrium
densities
\begin{eqnarray}\label{eq:surfacemodeslin}
n_{{\rm{eq}}}^{{\rm{TF}}}({x_{{\rm{in,out}}}}) =  -
\frac{{{F_{{\rm{in,out}}}}}}{U}{x_{{\rm{in,out}}}},
\end{eqnarray}
with $F_{\rm{in,out}}=-\nabla V_{\rm{bubble}}(R_{\rm{in,out}})$
and $x_{\rm{in,out}}$ the local variable pointing along the
direction of $F_{\rm{in,out}}$. This description of the condensate
is appropriate when the trapping potential $V_{\rm{bubble}}$
varies slowly over a distance
$\delta_{\rm{sm}}=[\hbar^2/(2m|F|)]^{1/3}$ from a condensate
boundary. Consequently, such linearization is not appropriate in
the very thin shell limit thus restricting
Eq.~(\ref{eq:surfacemodeslin}) to hollow BECs of nontrivial
thickness. In thin hollow condensates with thickness
$\delta_{\rm{t}}<2\delta_{\rm{sm}}$ surface modes confined to the
inner and the outer boundary overlap and hence cannot be treated
as strictly localized at either of these surfaces.

\begin{figure}[t]
\centering
\includegraphics[width=8.6cm]{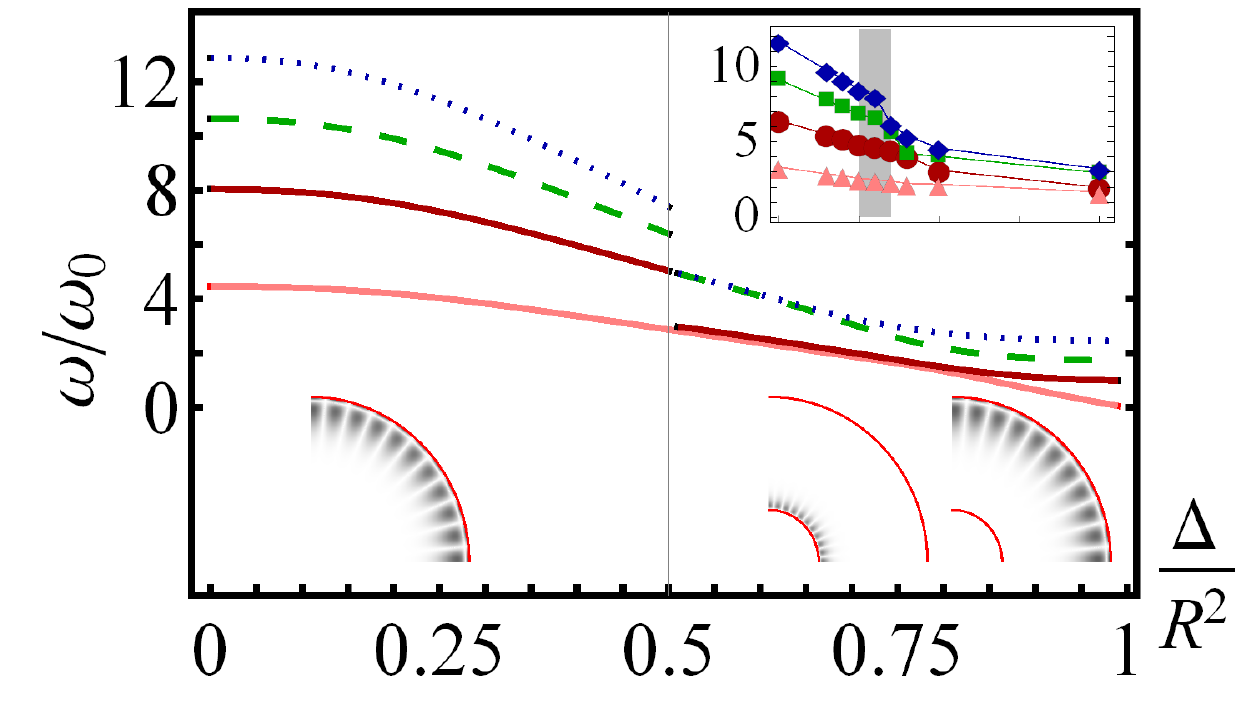}
\caption{(Color online) Oscillation frequencies of the $\ell=20$
collective modes for the lowest four radial index values
$\nu=0,1,2,3$ (curves from bottom to top, respectively) as the
system is tuned from a filled sphere $\Delta =0$ to a thin shell
$\Delta \sim R^2$ at a constant chemical potential. The line
through $\Delta=0.5 R^2$ (or the shaded region in the inset)
demarcates the filled and hollow regions and the schematics along
the bottom show the associated density deviations. Main: using the
Thomas-Fermi approximation for the equilibrium density, which
corresponds to sharp boundaries, and assuming the spatial extent
of the condensate does not change, i.e., $R$ is constant. Inset:
using equilibrium profiles given by numerical solution of the GP
equation (axes units are defined by the interaction energy
$UN=10^{4}$), which gives realistic soft boundaries, and assuming
that the number of atoms making up the condensate, $N$, is
constant.} \label{fig:Fig2}
\end{figure}

Surface mode frequencies for thick BEC shells can be derived from
the master equation, Eq.~(\ref{eq:delta_n_no_TF2}), using these
density profiles and the ansatz that the density distortions
exponentially decay into the bulk~\cite{AlKhawaja1999}.The
resultant frequencies are given by
$mS_l^2\omega^2_{\rm{in,out}}=(1+2\nu_{\rm{in,out}})F_{\rm{in,out}}q_{\rm{in,out}}$
where $\nu_{\rm{in,out}}$ indicates the number of nodes of the
collective mode confined to a particular condensate boundary in
the direction transverse to its surface. The wave-number
associated with each surface mode is given by
$q_{\rm{in,out}}=\ell/R_{\rm{in,out}}$. To be more precise, these
surface mode frequencies read
\begin{eqnarray}\label{eq:surfaceomega}
\omega^2_{\mathrm{in,out}}=\frac{\omega_0^2\ell(R^2-\Delta)}{\sqrt{(R^2-\Delta)^2/4+\Omega^2}}(2\nu_{\rm{in,out}}+1).
\end{eqnarray} We note that Eq.~(\ref{eq:surfaceomega}) is rather similar to the well-established
result for the surface modes of a fully filled spherical
condensate where
$\omega^2_{\rm{sp}}=\ell\omega_0^2(2\nu_{\rm{out}}+1)$ --- the
functional form and the $\ell$ dependence are the same up to
numerical factors accounting for the finite thickness of the
hollow shell BEC~\cite{AlKhawaja1999}. For the bubble trap,
$F_{\rm{in}}q_{\rm{in}}=F_{\rm{out}}q_{\rm{out}}$ which leads to a
degeneracy in the frequency of surface modes at the inner and
outer surfaces, as noted above. A simple explanation for this
degeneracy is that even though the inner surface is smaller in
area, its lower stiffness can support more oscillations per unit
distance (since $q_{\mathrm{in}}>q_{\mathrm{out}}$), bringing the
frequency of oscillations with $\nu_{\rm{in}}=\nu_{\rm{out}}$
nodes on the two surfaces into alignment. We note that increasing
the bubble trap parameter $\Delta$ decreases the thickness of the
condensate shell and leads to coupling between the inner and outer
boundary surface modes which can lift this degeneracy (as we will
show in Fig.~\ref{fig:Fig2}).

The restructuring of the surface mode spectrum following the
hollowing out transition thus presents a marked contrast to the
behavior of the same modes of a fully filled spherical BEC
discussed in the literature~\cite{AlKhawaja1999}. We remark that
inducing true surface modes corresponding to large $\ell$ values
presents an experimental challenge, and is yet to be achieved even
in the filled sphere case (although such modes have been
experimentally observed in classical fluids\cite{Brunet2011}). But
a comparison of the two cases highlights the difference between
the collective mode spectra of filled and hollow BECs in any
angular-momentum $\ell$ regime, including the two extreme limits:
$\ell=0$ and $\ell \gg 1$. We proceed to discuss the $\ell=0$
limit.

\section{Evolution of breathing modes}
Collective modes involving density distortions transverse to the
boundaries in a filled-sphere condensate correspond to spherically
symmetric breathing modes. As the system evolves to a hollow
shell, the density at the center of the system decreases, thus
creating a region of low stiffness where density oscillations tend
to localize and their frequency decreases. When the density in the
center of the system vanishes, the inner boundary is created and
the density deviations return to the bulk of the system. Thus, the
hollowing transition and the appearance of an inner boundary are
signaled by a prominent decrease in the transverse mode
frequencies. More rigorously, in Eq.~(\ref{eq:delta_n_no_TF2}),
the eigen-frequency receives two contributions: one proportional
to the equilibrium density, $n_{\rm{eq}}$, and the other to its
gradient, $\nabla n_{\rm{eq}}$. In general, $n_{\rm{eq}}$ is
smallest at the surface(s) of the BEC while $\nabla n_{\rm{eq}}$
is smallest at the location of extremum density. The only
situation in which both of these contributions can be small is at
the hollowing-out transition: at the center of the system, where
the inner boundary is emerging, both the condensate density and
its gradient simultaneously vanish. The hydrodynamic equation
therefore implies that the lowest frequency of oscillation for
breathing modes can be realized close to the center of the
condensate as it is starting to hollow. Consequently, the density
deviations for the collective modes concentrate near the nascent
inner surface and reach unprecedentedly low frequencies. The
spectrum of the breathing modes, therefore, displays a minimum in
frequency at the hollowing transition. Breathing modes exhibiting
this non-monotonic spectral property are among the most
experimentally accessible excitations of a three-dimensional
spherically symmetric BEC and thus a good candidate for an
observation of the effects of the hollowing-out transition.
Additionally, collective modes with low angular momentum values,
such as $\ell=1$ or $\ell=2$ exhibit similar frequency dip
features. In fact, in a realistic experimental system that may
lack perfect spherical symmetry, these low-$\ell$ collective modes
would be the most likely candidate of study.

\section{Numerical analyses for bubble-trap geometries}
Corroborating our heuristic arguments and simple derivations, we
now perform an in-depth numerical analysis for collective modes in
the bubble trap of Eq.~(\ref{eq:bubble}) by directly solving the
hydrodynamic equation of motion, Eq.~(\ref{eq:eigenproblem}) via a
finite-difference method~\cite{Smith1985}. We initially use the
Thomas-Fermi approximation for the equilibrium density
($n_{\rm{eq}}$); this approach focuses on salient features of the
topological hollowing transition by modeling a sharp inner
(Thomas-Fermi) boundary. In these calculations we hold the outer
radius of the condensate, $R$, fixed as $\Delta$ is changed. We
then go beyond the Thomas-Fermi approximation by solving the
time-independent Gross-Pitaevskii (GP) equation~\cite{Pethick2008}
for the equilibrium density, using an imaginary-time
algorithm~\cite{ChiofaloI2000}. Here, calculations are performed
with a fixed number of atoms $N$ in the condensate. This method
addresses the fate of the mode spectra in the physically realistic
case of soft condensate boundaries.

First addressing the surface modes, Fig.~\ref{fig:Fig2} shows the
evolution of the collective mode spectrum obtained by direct
numerical solution of Eq.~(\ref{eq:eigenproblem}) with $\ell=20$,
for a range of values of $\Delta$ in the bubble trap potential of
Eq.~(\ref{eq:bubble}), tuning from the fully filled sphere to thin
shell limits. The transition point where an inner boundary first
appears occurs at $R_{\rm{in}}=0$ or $\Delta/R^2=0.5$. As argued
on general grounds above, the transition from filled sphere to
hollow shell is indeed signaled by a reduction in frequency for a
given radial index and a sudden degeneracy between modes having
radial indices $2\nu$ and $2\nu+1$ (here, $\nu$ denotes the total
number of radial zeros for a given collective mode). Density
fluctuations on either side of the transition are schematically
represented along the bottom of Fig.~\ref{fig:Fig2}.

The results described above model a sudden change from the filled
sphere to the shell topology through the appearance of the sharp
inner boundary in the Thomas-Fermi approximation. In contrast, the
inset to Fig.~\ref{fig:Fig2} shows the result of solving the
hydrodynamic equation for the $\ell=20$ collective modes using the
numerically exact equilibrium density, $n_{\rm{eq}}$, obtained by
solving the time-independent GP equation (with an interaction
strength of $NU=10^4$). This more realistic modeling of the
gradually emerging inner surface yields a smearing of the sharp
frequency discontinuity found in the Thomas-Fermi approximation as
marked by the shaded region in the inset of the figure.
Nevertheless, the frequency spectrum of the collective modes still
displays a notable jump and near-degeneracy of modes as the inner
boundary emerges.

\begin{figure}[t]
\centering
\includegraphics[width=8.6cm]{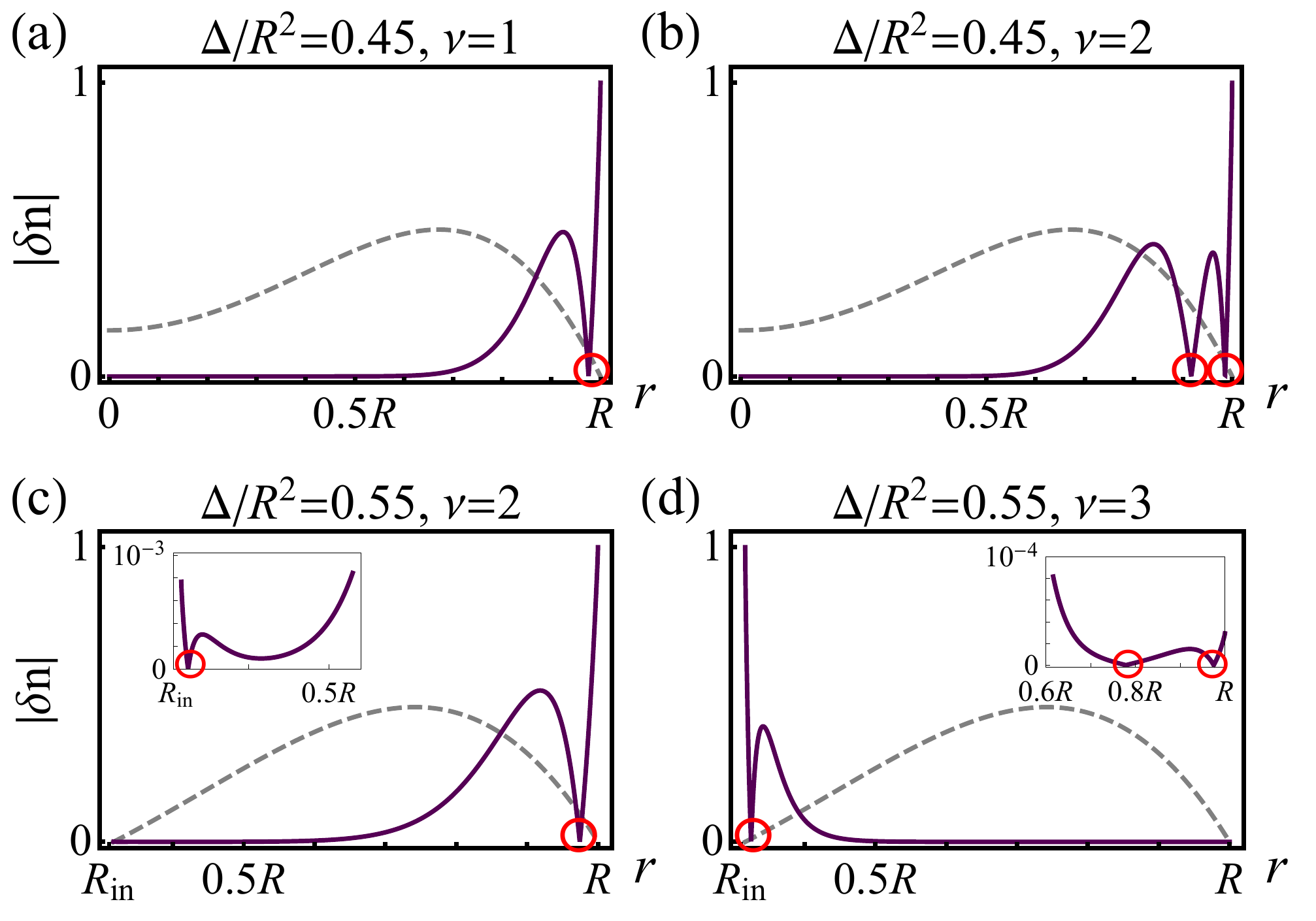}
\caption{(Color online) Normalized density deviation profiles
$|\delta n (r)|$ along the radial direction of surface collective
modes  ($\ell = 20$) in filled (a,b) and hollow (c,d) condensates
with $\Delta/R^2 = 0.45$ and 0.55, respectively, for varying
number of radial nodes $\nu$, as denoted. The density deviation
profile and radial node (circled) structure of $\nu=1$ in the
filled case resembles that of $\nu=2$ and $\nu=3$ in the hollow
case close to the outer and inner boundaries, respectively. In the
barely hollow situation shown, the centrifugal barrier given by
$\ell (\ell+1) n_{\rm{eq}}$ in Eq.~(\ref{eq:eigenproblem}) -- and
represented by the dotted line (in arbitrary units) -- is lowered
so as to  distribute the remaining nodes of the $\nu=2$ and
$\nu=3$ to the other surface as small oscillations [insets in
(c,d)].} \label{fig:Fig3}
\end{figure}

To corroborate our description of the frequency restructuring
across the hollowing transition, in Fig.~\ref{fig:Fig3} we present
the density deviations and radial node redistribution in the
illustrative example of surface modes having $\ell=20$ and
$\nu=1,2$ and $3$ on either side of the topological transition.
Comparing Figs.~\ref{fig:Fig3} (a), (c) and (d), it is clear that
after the emergence of the condensate's inner boundary, both modes
with total number of radial nodes $\nu=2$ and $\nu=3$ bear
similarity, in density deviations, to the mode with $\nu=1$ radial
nodes before this boundary surface was available. Additional nodes
for $\nu=2$ and $\nu=3$ collective modes are present but appear as
radial oscillations of small amplitude on a boundary surface
opposite to the one with the most prominent nodal structure. After
the hollowing change, the overall oscillation frequencies of the
$\nu=2$ and $\nu=3$ collective modes are accordingly dominated by
a single node associated with high amplitude radial motion. As
this motion has the highest energetic cost, these frequencies are
nearly equivalent to the oscillation frequency of the collective
mode with a single radial node in the fully filled spherical
condensate. We emphasize that this is a feature that only occurs
once the system is hollow, as demonstrated by a comparison of
Figs.~\ref{fig:Fig3} (a) and (b)---collective modes with $\nu=1$
and $\nu=2$ radial nodes have rather different radial profiles for
a fully filled spherical BEC. Extrapolating to more general
surface modes, this redistribution of nodes explains the jump in
the frequency spectrum shown in Fig.~\ref{fig:Fig2} where
collective modes with total $2\nu$ and $2\nu+1$ radial nodes after
the hollowing transition smoothly continue the spectrum of the
$\nu$ surface mode of the BEC prior to hollowing.

\begin{figure}[t]
\centering
\includegraphics[width=8.6cm]{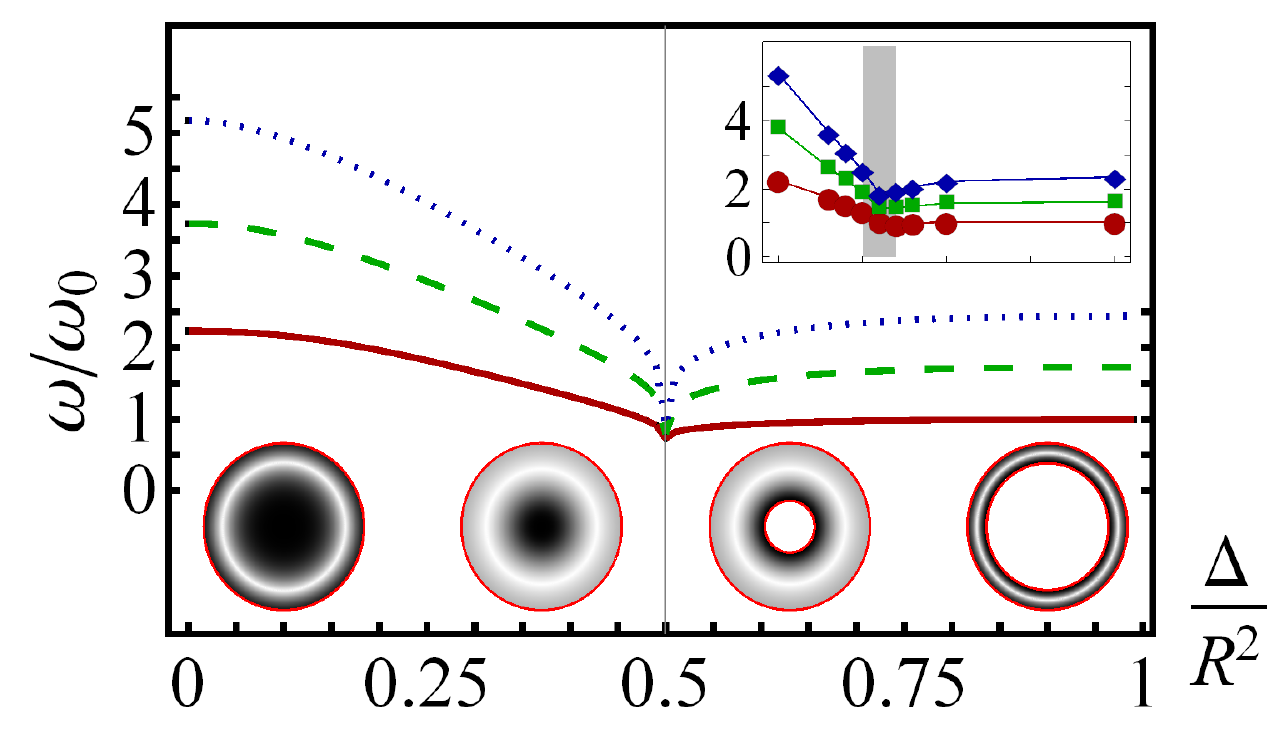}
\caption{(Color online) Oscillation frequencies of the lowest
three breathing ($\ell=0$) collective modes $\nu=1,2,3$ (curves
from bottom to top) as $\Delta$ is varied from the filled-sphere
to the thin-shell limit at a constant chemical potential, in the
Thomas-Fermi approximation with constant $R$ (main) and using the
numerically exact equilibrium densities with constant $N$ (inset).
The line through $\Delta=0.5 R^2$ (or the shaded region in the
inset) demarcates the filled and hollow regions and the schematics
along the bottom show the associated density deviations, for (from
left to right) $\Delta/R^2 =0$, $0.45$, $0.55$, and $0.8$.}
\label{fig:Fig4}
\end{figure}

In Fig.~\ref{fig:Fig4}, to exhibit the behavior of the breathing
modes, we show the results of the direct numerical solution of
Eq.~(\ref{eq:eigenproblem}) for the case $\ell=0$ as the parameter
$\Delta$ is varied. We note the sharp depression of the
frequencies at the transition point and their otherwise smooth
evolution to the solid-sphere and thin-shell limits at
$\Delta/R^2=0$ and $1$, respectively. Additionally, we note that
the density deviation for these spherically symmetric modes, as
shown along the bottom of Fig.~\ref{fig:Fig4}, displays
confinement to the inner boundary at the transition from filled to
hollow. As with the surface modes, we expect this sudden
transition to be replaced by a smoother crossover in a realistic
system in which the condensate's boundaries are not sharp. In the
inset to Fig.~\ref{fig:Fig4}, we show the result of solving the
hydrodynamic equation for the $\ell=0$ collective modes using the
numerical solution to the time-independent GP equation for the
equilibrium density, $n_{\rm{eq}}$. While in this more realistic
situation the transition is gradual, as indicated by the shaded
region in the inset of Fig.~\ref{fig:Fig3}, the characteristic
decrease in frequency from the filled sphere through the
transition point clearly persists, indicating that the hollowing
at the center of the condensate can be observed through the
spectrum of the system's breathing modes. Additionally, since the
decrease in central density here occurs less sharply than when the
Thomas-Fermi approximation is used, the frequency curves in the
inset of Fig.~\ref{fig:Fig4} decrease to distinct values at the
hollowing transition, in contrast to their near-degeneracy in the
Thomas-Fermi approximation.

To summarize, we have explored the physics of a BEC undergoing a
topological change from a filled sphere to a hollow shell and
shown that the collective modes are powerful indicators of such a
transition. In particular, spherically symmetric breathing modes
show a crossover between the fully filled and thin shell limits at
the hollowing-out point where the associated frequency spectrum
exhibits a tell-tale dip. High-angular-momentum surface modes show
marked sensitivity to the appearance of a new boundary as well.

We now turn to the analogous lower-dimensional BEC geometry of a
disk hollowing out into an annulus, effectively represented by a
toroidal geometry. Both structures have been well-studied
theoretically~\cite{Stringari2006, Perrin2012} and
experimentally~\cite{Gupta2005, Ramanathan2011}. The transition
regime between the two topologies has yet to be studied and is
potentially experimentally more tracatable than the
three-dimensional spherical case. We have performed preliminary
collective mode spectral analyses for the disk-annulus system
using the same techniques as those presented here. Our main
results for the spherical case hold equally well for the two
dimensional system, given the common physics (vanishing of central
density at the hollowing out transition and so on). Specifically,
the breathing modes show a dip and surface modes a reconfiguration
of radial nodes in their frequency spectra around the topological
transition from a disk to an annulus. While the shell and the
annulus topologies host similar collective modes when their
thickness is small~\cite{Stringari2006}, the collective mode
frequencies in the overall hollow regimes differ. Moreover, the
collective mode spectra cannot distinguish the different
topological measures of the BEC's shape such as the fundamental
group for the disk-annulus transition (determined by whether or
not a loop can be shrunk to a point) or the second homotopy group
for the sphere-shell transition (determined by whether a spherical
surface can be shrunk to a point).  However, we note that a
fundamental difference exists between the two cases, making the
study of collective mode spectra more pressing in the
three-dimensional case: in two dimensions, the presence of a
hollow inner region can be imaged (and visualized) directly by
absorption imaging techniques. However, for the spherical
geometry, direct imaging of the hollow region is obstructed by the
presence of a surrounding condensate in all three dimensions,
rendering collective mode spectra a powerful, non-destructive
probe of the hollowing transition.

Finally, the CAL trap experiment is expected to pioneer the
realization of condensate shells thus adding to the growing
interest in BECs in microgravity conditions in space~\cite{MAIUS}
and Earth-based environments~\cite{ZARM}. The challenge has been
that under typical terrestrial experimental conditions,
gravitational sag causes heavy depletion at the shell's apex and a
pooling of atoms at its bottom. Excluding gravitational effects,
we estimate that for a BEC of $\sim 10^5$ $^{87}$Rb atoms forming
a cloud of $10~\mu$m in a harmonic trap of bare frequency $500$
Hz, in the transition from the filled sphere to the thin shell
limit the lowest breathing (surface) mode would evolve from about
1 (4) kHz to 0.5 kHz, which are in an accessible regime and in a
large enough range to probe our predictions. We predict the
decrease in the collective mode frequency at the hollowing point,
compared to the oscillation frequency of the same mode in the
fully filled spherical BEC to be rather prominent, on the order of
$50\%$ or more, for all low-lying breathing (and low-$\ell$)
modes. Somewhat higher modes, such as $\nu=3$, are suitable for
for experimental detection of the diminishing of collective mode
frequency at the hollowing point compared to the thin-shell limit
as well. For instance, the collective mode with $\nu=3$ shows a
$20\%$ change between these two regimes, which makes it a good
candidate for a full observation of the non-monotonicity of the
collective mode frequency spectrum at the hollowing point. Further
work on BEC shells from collective mode behavior, expansion and
time-of-flight, and the nature of vortices in this new geometry is
in order. These studies are relevant to situations ranging from
the microscopic scale of the CAL trap experiments to the proposed
existence of BEC shells in stellar objects and would deepen our
insight into topological changes in quantum systems.

\acknowledgments We thank Nathan Lundblad and Michael Stone for
illuminating discussions. KS acknowledges support by ARO
(W911NF-12-1-0334), AFOSR (FA9550-13-1-0045), NSF (PHY-1505496),
and Texas Advanced Computing Center (TACC). CL acknowledges
support by the National Science Foundation under award
DMR-1243574. KP, SV, and CL acknowledge support by NASA (SUB JPL
1553869 and 1553885). CL and SV thank the KITP for hospitality.


\begin{thebibliography}{0}


\bibitem{Thouless1973Haldane1983}J. M. Kosterlitz and D. J. Thouless, Ordering, metastability and phase transitions in two-dimensional systems,
\href{https://doi.org/10.1088/0022-3719/6/7/010}{J. Phys. C: Solid
State Phys. \textbf{6}, 1181 (1973)}; F. D. M. Haldane, Nonlinear
Field Theory of Large-Spin Heisenberg Antiferromagnets:
Semiclassically Quantized Solitons of the One-Dimensional
Easy-Axis N\'eel State,
\href{https://doi.org/10.1103/PhysRevLett.50.1153}{Phys. Rev.
Lett. \textbf{50}, 1153 (1983)}.

\bibitem{Hasan2010Qi2011Chiu2016}M. Z. Hasan and C. L. Kane, Topological Insulators,
\href{https://doi.org/10.1103/RevModPhys.82.3045}{Rev. Mod. Phys.
\textbf{82}, 3045 (2010)}; X.-L. Qi and S.-C. Zhang, Topological
Insulators and Superconductors,
\href{https://doi.org/10.1103/RevModPhys.83.1057}{Rev. Mod. Phys.
\textbf{83}, 1057 (2011)}; C.-K. Chiu, J. C. Y. Teo, A. P.
Schnyder, and S. Ryu, Classification of topological quantum matter
with symmetries,
\href{https://doi.org/10.1103/RevModPhys.88.035005}{Rev. Mod.
Phys. \textbf{88}, 035005 (2016)}.


\bibitem{Atala2013} M. Atala, M. Aidelsbrger, J. T. Barreiro, D. Abanin, T. Kitagawa, E. Demler and I. Bloch, Direct measurements of the Zak phase in topological Bloch bands,
\href{https://doi.org/10.1038/nphys2790}{Nat. Phys. \textbf{9},
795 (2013)}.

\bibitem{Delplace2011} P. Delplace, D. Ullmo and G.Montambaux, Zak phase and the existence of edge state in graphene,
\href{https://doi.org/10.1103/PhysRevB.84.195452}{Phys. Rev. B
\textbf{84}, 195452 (2011)}.

\bibitem{Price2013}H. M. Price and N. R. Cooper, Effects of Berry Curvature on the Collective Modes of Ultracold Gases
\href{https://doi.org/10.1103/PhysRevLett.111.220407}{Phys. Rev.
Lett. \textbf{111}, 220407 (2013)}.


\bibitem{Gupta2005} S. Gupta, K. W. Murch, K. L. Moore, T. P. Purdy, and D. M. Stamper-Kurn, Bose-Einstein Condensation in a Circular Waveguide,
\href{https://doi.org/10.1103/PhysRevLett.95.143201}{Phys. Rev.
Lett. \textbf{95}, 143201 (2005)}.

\bibitem{Ramanathan2011} A. Ramanathan, K. C. Wright, S. R. Muniz, M. Zelan, W. T. Hill, III, C. J. Lobb, K. Helmerson, W. D. Phillips, and G. K. Campbell, Superflow in a Toroidal Bose-Einstein Condensate: An Atom Circuit with a Tunable Weak Link,
\href{https://doi.org/10.1103/PhysRevLett.106.130401}{Phys. Rev.
Lett. \textbf{106}, 130401 (2011)}.


\bibitem{CAL}\href{http://coldatomlab.jpl.nasa.gov}{http://coldatomlab.jpl.nasa.gov}

\bibitem{Lundblad}N. Lundblad, private communication (2015).

\bibitem{Hollow}N. Lundblad, T. Jarvis, D. Paseltiner, and C. Lannert, Progress toward studies of bubble-geometry Bose-Einstein condensates in microgravity with a ground-based prototype of NASA CAL,
\href{http://meetings.aps.org/Meeting/DAMOP16/Session/K1.119}{DAMOP Meeting, K1.00119 (2016)}.


\bibitem{Zobay2001}O. Zobay and B. M. Garraway, Two-Dimensional Atom Trapping in Field-Induced Adiabatic Potentials,
\href{https://doi.org/10.1103/PhysRevLett.86.1195}{Phys. Rev.
Lett. \textbf{86}, 1195 (2001)}; Atom trapping and two-dimensional
Bose-Einstein condensates in field-induced adiabatic potentials,
\href{https://doi.org/10.1103/PhysRevA.69.023605}{Phys. Rev. A
\textbf{69}, 023605 (2004)}.


\bibitem{Gorlitz2001}
A. Gorlitz, J.M. Vogels, A. E Leanhardt, C. Raman, T. L.
Gustavson, J. R. Abo-Shaeer, A. P. Chikkatur, S. Gupta, S. Inouye,
T. Rosenband, and W. Ketterle, Realization of Bose-Einstein
Condensates in Lower Dimensions,
\href{https://doi.org/10.1103/PhysRevLett.87.130402}{Phys. Rev.
Lett. \textbf{87}, 130402 (2001)}.

\bibitem{Greiner2001} M. Greiner, I. Bloch, O. Mandel, T. W. Hansch, and T. Esslinger, Exploring Phase Coherence in a 2D Lattice of Bose-Einstein Condensates,
\href{https://doi.org/10.1103/PhysRevLett.87.160405}{Phys. Rev.
Lett. \textbf{87}, 160405 (2001)}.

\bibitem{Dettemer2001} S. Dettemer, D. Hellweg, P. Ryytty, J. J. Arlt, W. Ermer, K. Sengstock, D. S. Petrov, G. V. Shlyapnikov, H. Kreutzmann, L. Santos, and M. Lewenstein, Observation of Phase Fluctuations in Elongated Bose-Einstein Condensates,
\href{https://doi.org/10.1103/PhysRevLett.87.160406}{Phys. Rev.
Lett. \textbf{87}, 160406 (2001)}.

\bibitem{Hechenblaikner2005} G. Hechenblaikner, J. M. Krueger, and C. J. Foot, Properties of quasi-two-dimensional Condensates in Highly Anisotropic Traps,
\href{https://doi.org/10.1103/PhysRevA.71.013604}{Phys. Rev. A
\textbf{71}, 013604 (2005)}.

\bibitem{Smith2005} N. L. Smith, W. H. Heathcote, G. Hechenblaikner, E. Nugent, and C. J. Foot, Quasi-2D Confinement of a BEC in a Combined Optical and Magnetic Potential,
\href{https://doi.org/10.1088/0953-4075/38/3/007}{J. Phys. B
\textbf{38}, 223 (2005)}.


\bibitem{Gaunt2013} A. L. Gaunt, T. F. Schmidutz, I. Gotlibovych, R. P. Smith, and Z. Hadzibabic, Bose-Einstein Condensation of Atoms in a Uniform Potential,
\href{https://doi.org/10.1103/PhysRevLett.110.200406}{Phys. Rev.
Lett. \textbf{110}, 200406 (2013)}.


\bibitem{Smerzi1997} A. Smerzi, S. Fantoni, S. Giovanazzi, and S. R. Shenoy, Quantum Coherent Atomic Tunneling between Two Trapped Bose-Einstein Condensates,
\href{https://doi.org/10.1103/PhysRevLett.79.4950}{Phys. Rev.
Lett. \textbf{79}, 4950 (1997)}.

\bibitem{Shin2004} Y. Shin, M. Saba, T. A. Pasquini, W. Ketterle, D. E. Pritchard, and A. E. Leanhardt, Atom Interferometry with Bose-Einstein Condensates in a Double-Well Potential,
\href{https://doi.org/10.1103/PhysRevLett.92.050405}{Phys. Rev.
Lett. \textbf{92}, 050405 (2004)}.

\bibitem{Hofferberth2006} S. Hofferberth, I. Lesanovsky, B. Fischer, J. Verdu, and J. Schmiedmayer, Radiofrequency-dressed-state Potentials for Neutral Atoms,
\href{https://doi.org/10.1038/nphys420}{Nat. Phys. \textbf{2}, 710
(2006)}.


\bibitem{Colombe2004}Y. Colombe, E. Knyazchyan, O. Morizot, B. Mercier, V. Lorent, and H. Perrin, Ultracold atoms confined in rf-induced two-dimensional trapping potentials,
\href{https://doi.org/10.1209/epl/i2004-10095-7}{EPL \textbf{67},
593 (2004)}.

\bibitem{Merloti2013}K. Merloti, R. Dubessy, L. Longchampbon, A. Perrin, P.-E. Pottie, V. Lorent, and H. Perrin, A two-dimensional quantum gas in a magnetic trap,
\href{https://doi.org/10.1088/1367-2630/15/3/033007}{New J. Phys.
\textbf{15}, 033007 (2013)}.


\bibitem{ZARM}\href{http://www.zarm.uni-bremen.de/drop-tower.html}{http://www.zarm.uni-bremen.de/drop-tower.html}


\bibitem{Wang94} T. G. Wang, A. V. Anilkumar, C. P. Lee and K. C. Lin, Core-centering of Compound Drops in Capillary Oscillations: Observations on USML-1 Experimets in Space,
\href{https://doi.org/10.1006/jcis.1994.1201}{J. Colloid Interface
Sci \textbf{165}, 1 (1994)}.


\bibitem{Molmer1998}K. M{\o}lmer, Bose Condensates and Fermi Gases at Zero Temperature,
\href{https://doi.org/10.1103/PhysRevLett.80.1804}{Phys. Rev.
Lett. \textbf{80}, 1804 (1998)}.

\bibitem{Ospelkaus2006}S. Ospelkaus, C. Ospelkaus, L. Humbert, K. Sengstock, and K. Bongs, Tuning of Heteronuclear Interactions in a Degenerate Fermi-Bose Mixture,
\href{https://doi.org/10.1103/PhysRevLett.97.120403}{Phys. Rev. Lett. \textbf{97}, 120403 (2006)}.

\bibitem{Schaeybroeck2009}B. V. Schaeybroeck and A. Lazarides, Trapped phase-segregated Bose-Fermi mixtures and their collective excitations,
\href{https://doi.org/10.1103/PhysRevA.79.033618}{Phys. Rev. A
\textbf{79}, 033618 (2009)}.


\bibitem{Batrouni2002}G. Batrouni, V. Rousseau, R. Scalettar, M. Rigol, A. Muramatsu, P. Denteneer, and M. Troyer, Mott Domains of Bosons Confined on Optical Lattices,
\href{https://doi.org/10.1103/PhysRevLett.89.117203}{Phys. Rev.
Lett. \textbf{89} 117203 (2002)}.

\bibitem{DeMarco2005}B. DeMarco, C. Lannert, S. Vishveshwara, and T.-C. Wei, Structure and stability of Mott-insulator shells of bosons trapped in an optical lattice,
\href{https://doi.org/10.1103/PhysRevA.71.063601}{Phys. Rev. A
\textbf{71}, 063601 (2005)}.
\bibitem{Campbell2006}G. K. Campbell, J. Mun, M. Boyd, P. Medley, A. E. Leanhardt, L. G. Marcassa, D. E. Pritchard, and W. Ketterle, Imaging the Mott Insulator Shells by Using Atomic Clock Shifts,
\href{https://doi.org/10.1126/science.1130365}{Science
\textbf{313}, 649 (2006)}.

\bibitem{Barankov2007Sun2009}R. A. Barankov, C. Lannert, and S. Vishveshwara, Coexistence of superfluid and Mott phases of lattice bosons,
\href{https://doi.org/10.1103/PhysRevA.75.063622}{Phys. Rev. A
\textbf{75}, 063622 (2007)}; K. Sun, C. Lannert, and S.
Vishveshwara, Probing condensate order in deep optical lattices,
\href{https://doi.org/10.1103/PhysRevA.79.043422}{Phys. Rev. A
\textbf{79}, 043422 (2009)}.


\bibitem{Weber2004}F. Weber, Strange quark matter and compact stars,
\href{https://doi.org/10.1016/j.ppnp.2004.07.001}{Prog. Part.
Nucl. Phys. \textbf{54}, 193 (2004)}.

\bibitem{Pethick2015}C. J. Pethick, T. Schaefer, and A. Schwenk, Bose-Einstein condensates in neutron stars,
\href{https://arxiv.org/abs/1507.05839}{arXiv:1507.05839}.


\bibitem{Kac1966}M. Kac, Can One Hear the Shape of a Drum?,
\href{https://doi.org/10.2307/2313748}{Am. Math. Monthly
\textbf{73}, 1 (1966)}.

\bibitem{Giraud2010}O. Giraud and K. Thas, Hearing shapes of drums: Mathematical and physical aspects of isospectrality,
\href{https://doi.org/10.1103/RevModPhys.82.2213}{Rev. Mod. Phys. \textbf{82}, 2213 (2010)}.


\bibitem{Stringari1996}S. Stringari, Collective Excitations of a Trapped Bose-Condensed Gas,
\href{https://doi.org/10.1103/PhysRevLett.77.2360}{Phys. Rev.
Lett. \textbf{77}, 2360 (1996)}.

\bibitem{Jin1996}D. S. Jin, J. R. Ensher, M. R. Matthews, C. E. Wieman, and E. A. Cornell, Collective Excitations of a Bose-Einstein Condensate in a Dilute Gas,
\href{https://doi.org/10.1103/PhysRevLett.77.420}{Phys. Rev. Lett.
\textbf{77}, 420 (1996)}.

\bibitem{Mewes1996}M.-O. Mewes, M. R. Andrews, N. J. van Druten, D. M. Kurn, D. S. Durfee, C. G. Townsend, and W. Ketterle, Collective Excitations of a Bose-Einstein Condensate in a Magnetic Trap,
\href{https://doi.org/10.1103/PhysRevLett.77.988}{Phys. Rev. Lett.
\textbf{77}, 988 (1996)}.

\bibitem{Edwards1996}M. Edwards, P. A. Ruprecht, K. Burnett, R. J. Dodd, and C. W. Clark, Collective Excitations of Atomic Bose-Einstein Condensates,
\href{https://doi.org/10.1103/PhysRevLett.77.1671}{Phys. Rev.
Lett. \textbf{77}, 1671 (1996)}.

\bibitem{Castin1996}Y. Castin and R. Dum, Bose-Einstein Condensates in Time Dependent Traps,
\href{https://doi.org/10.1103/PhysRevLett.77.5315}{Phys. Rev.
Lett. \textbf{77}, 5315 (1996)}.

\bibitem{Stamper-Kurn1998}D. M. Stamper-Kurn, H.-J. Miesner, S. Inouye, M. R. Andrews, and W. Ketterle, Collisionless and Hydrodynamic Excitations of a Bose-Einstein Condensate,
\href{https://doi.org/10.1103/PhysRevLett.81.500}{Phys. Rev. Lett.
\textbf{81}, 500 (1998)}.

\bibitem{Chevy2002}F. Chevy, V. Bretin, P. Rosenbusch, K. W. Madison, and J. Dalibard, Transverse Breathing Mode of an Elongated Bose-Einstein Condensate,
\href{https://doi.org/10.1103/PhysRevLett.88.250402}{Phys. Rev.
Lett. \textbf{88}, 250402 (2002)}.

\bibitem{Fort2003}C. Fort, F. S. Cataliotti, L. Fallani, F. Ferlaino, P. Maddaloni, and M. Inguscio, Collective Excitations of a Trapped Bose-Einstein Condensate in the Presence of a 1D Optical Lattice,
\href{https://doi.org/10.1103/PhysRevLett.90.140405}{Phys. Rev.
Lett. \textbf{90}, 140405 (2003)}.

\bibitem{Yang2009}L. Yang, X.-R. Wang, Ke Li, X.-Z. Tan, H.-W. Xiong, and B.-L. Lu, Low-Energy Collective Excitation of Bose-Einstein Condensates in an Anisotropic Magnetic Trap,
\href{https://doi.org/10.1088/0256-307X/26/7/076701}{Chin. Phys.
Lett. \textbf{26}, 076701 (2009)}.

\bibitem{Haller2009}E. Haller, M. Gustavsson, M. J. Mark, J. G. Danzl, R. Hart, G. Pupillo, and H.-C. N\"agerl, Realization of an Excited, Strongly Correlated Quantum Gas Phase,
\href{https://doi.org/10.1126/science.1175850}{Science
\textbf{325}, 1224 (2009)}.

\bibitem{Pollack2010}S. E. Pollack, D. Dries, R. G. Hulet, K. M. F. Magalh\~aes, E. A. L. Henn, E. R. F. Ramos, M. A. Caracanhas, and V. S. Bagnato, Collective excitation of a Bose-Einstein condensate by modulation of the atomic scattering length,
\href{https://doi.org/10.1103/PhysRevA.81.053627}{Phys. Rev. A
\textbf{81}, 053627 (2010)}.

\bibitem{Kuwamoto2012}T. Kuwamoto and T. Hirano, Collective Excitation of Bose-Einstein Condensates Induced by Evaporative Cooling,
\href{https://doi.org/10.1143/JPSJ.81.074002}{J. Phys. Soc. Jpn.
\textbf{81}, 074002 (2012)}.


\bibitem{Lighthill1978}J. Lighthill \emph{Waves in Fluids}, (Cambridge University Press, Cambridge, 1978).


\bibitem{Lannert2007} C. Lannert, T.-C. Wei and S. Vishvehswara,
Dynamics of condensate shells: Collective modes and expansion,
\href{https://doi.org/10.1103/PhysRevA.75.013611}{Phys. Rev. A
\textbf{75}, 013611, (2007)}.


\bibitem{Stringari2006} M. Cozzini, B. Jackson, and S. Stringari, Vortex Signatures in Annular Bose-Einstein Condensates,
\href{https://doi.org/10.1103/PhysRevA.73.013603}{Phys. Rev. A
\textbf{73}, 013603 (2006)}.

\bibitem{Perrin2012} R. Dubessy, T. Liennard, P. Pedri, and H. Perrin, Critical Rotation of an Annular Superfluid Bose-Einstein Condensate,
\href{https://doi.org/10.1103/PhysRevA.86.011602}{Phys. Rev. A
\textbf{86}, 011602(R) (2012)}.


\bibitem{AlKhawaja1999} U. Al Khawaja, C. J. Pethick, and H. Smith, {Surface of a Bose-Einstein condensed atomic cloud},
\href{https://doi.org/10.1103/PhysRevA.60.1507}{Phys. Rev. A
\textbf{60}, 1507 (1999).}


\bibitem{Lobser2015Straatsma2016}D. S. Lobser, A. E. S. Barentine, E. A. Cornell, and H. J. Lewandowski, Observation of a persistent non-equilibrium state in cold atoms,
\href{https://doi.org/10.1038/nphys3491}{Nat. Phys. \textbf{11},
1009 (2015)}; C. J. E. Straatsma, V. E. Colussi, M. J. Davis, D.
S. Lobser, M. J. Holland, D. Z. Anderson, H. J. Lewandowski, and
E. A. Cornell, {Collapse and revival of the monopole mode of a
degenerate Bose gas in an isotropic harmonic trap},
\href{https://doi.org/10.1103/PhysRevA.94.043640}{Phys. Rev. A
\textbf{94}, 043640 (2016)}.


\bibitem{Pethick2008}C. J. Pethick and H. Smith, \emph{Bose-Einstein Condensation in Dilute Gases}, 2st, ed. (Cambridge University Press, Cambridge, 2008).


\bibitem{Brunet2011} P. Brunet and J. H. Snoeijer, Star-drops formed by periodic excitation and on an air cushion -- A shot review,
\href{https://doi.org/10.1140/epjst/e2011-01375-5}{EPJ ST
\textbf{192}, 1 (2011)}


\bibitem{Smith1985}G. D. Smith, \emph{Numerical solution of partial differential equations: finite difference methods} (Oxford university press, 1985).


\bibitem{ChiofaloI2000}M. L. Chiofalo, S. Succi, and M. P. Tosi, Ground state of trapped interacting Bose-Einstein condensates by an explicit imaginary-time algorithm,
\href{https://doi.org/10.1103/PhysRevE.62.7438}{Phys. Rev. E
\textbf{62}, 7438 (2000)}.


\bibitem{MAIUS}\href{https://www.zarm.uni-bremen.de/en/research/space-technologies/fundamental-physics-tests-in-space/projects/quantus-iv-maius.html}{https://www.zarm.uni-bremen.de/en.html}

\end{thebibliography}
\end{document}